\documentclass[%
 reprint,
 amsmath,amssymb,
 aps,
prl,
floatfix]{revtex4-2}

\usepackage{soul,color}
\soulregister{\cite}{7}
\soulregister{\ref}{7}
\soulregister{\pageref}{7}

\usepackage[english]{babel}
\usepackage{graphicx}
\usepackage{dcolumn}
\usepackage{bm}
\usepackage{physics}
\usepackage{float}
\usepackage{hyperref}

\begin{document}

\preprint{APS/123-QED}

\title{The origin of KPZ-scaling in arrays of polariton condensates}

\author{Denis Novokreschenov}
\email{novokreshchenov.ds@phystech.edu}
\affiliation{%
 Abrikosov Center for Theoretical Physics, Moscow Center for Advanced Studies, Kulakova str. 20, Moscow, Russia
}%
\affiliation{Russian Quantum Center, Skolkovo, Moscow, 121205, Russia}

\author{Alexey Kavokin}
\affiliation{%
 Abrikosov Center for Theoretical Physics, Moscow Center for Advanced Studies, Kulakova str. 20, Moscow, Russia
}%
\affiliation{Russian Quantum Center, Skolkovo, Moscow, 121205, Russia}
\affiliation{Department of Physics, St. Petersburg State University, University Embankment, 7/9, St. Petersburg, 199034, Russia}
\affiliation{School of Science, Westlake University, 18 Shilongshan Road, Hangzhou 310024, Zhejiang Province, China}

\date{\today}

\begin{abstract}
This work investigates the origin of Kardar–Parisi–Zhang (KPZ) scaling in the phase dynamics of one-dimensional and two-dimensional polariton condensates. We demonstrate that the key mechanism leading to the observed power laws for the first-order correlation function $g^{(1)}$ is the fluctuation of the population of Goldstone modes, which arise due to the spontaneous breaking of $U(1)$ symmetry. Numerical simulations and analytical theory confirm that the critical exponents describing the KPZ universality class directly follow from the dynamics of Goldstone excitations. Our results establish a direct connection between the microscopic parameters of arrays of exciton-polariton condensates and the coherent properties of the light they emit.
\end{abstract}

\maketitle



\textit{Introduction.} The Kardar-Parisi-Zhang (KPZ) equation, originally formulated to describe the stochastic growth of interfaces~\cite{KPZ}, defines a broad universality class observed in diverse systems ranging from bacterial colony growth~\cite{KPZ_bio} to turbulent liquid crystals~\cite{KPZ_liquid_crystals}. In recent years, this universal scaling behavior has been unexpectedly discovered in the phase dynamics of spatially extended non-equilibrium quantum fluids, particularly in exciton-polariton condensates~\cite{KPZ_Fontaine, KPZ_Hofling}. These hybrid light-matter systems, formed in semiconductor microcavities~\cite{Weisbuch}, exhibit spontaneous coherence and non-equilibrium Bose-Einstein condensation under optical pumping~\cite{Kasprzak, Yamamoto_RMP, Yamamoto_Science, West}. A key feature of such condensates is the spontaneous breaking of the global $U(1)$ symmetry, which gives rise to massless Nambu-Goldstone modes—phase fluctuations of the condensate order parameter~\cite{Sanvitto_Goldstone, Wouters_Carusotto_Goldstone, Hsieh_Goldstone, Utsunomiya_NaturePhysics}.

The connection between phase dynamics and the KPZ universality class has been established using driven-dissipative Gross-Pitaevskii equation (GPE) with white noise term which is added phenomenologically~\cite{KPZ_Deligiannis_PRR, Deligiannis_EPL, KPZ_Novokreschenov, KPZ_Kulkarni, Eastham_PRR, Eastham_PRB, KPZ_Szymanska, Diehl_KPZ, Diehl_review}. But its fundamental microscopic explanation rooted in the specific properties of polaritonic systems has remained unexplained. In particular, the origin of the characteristic power-law decay of the first-order coherence function $g^{(1)}$, with its distinct critical exponents in one and two dimensions, calls for a derivation from the first principles that links the microscopic parameters — such as polariton-polariton interaction strength, energy dispersion, and effective temperature — to the emergent KPZ scaling.

\begin{figure}[h!]
    \centering
    \includegraphics[width=0.99\linewidth]{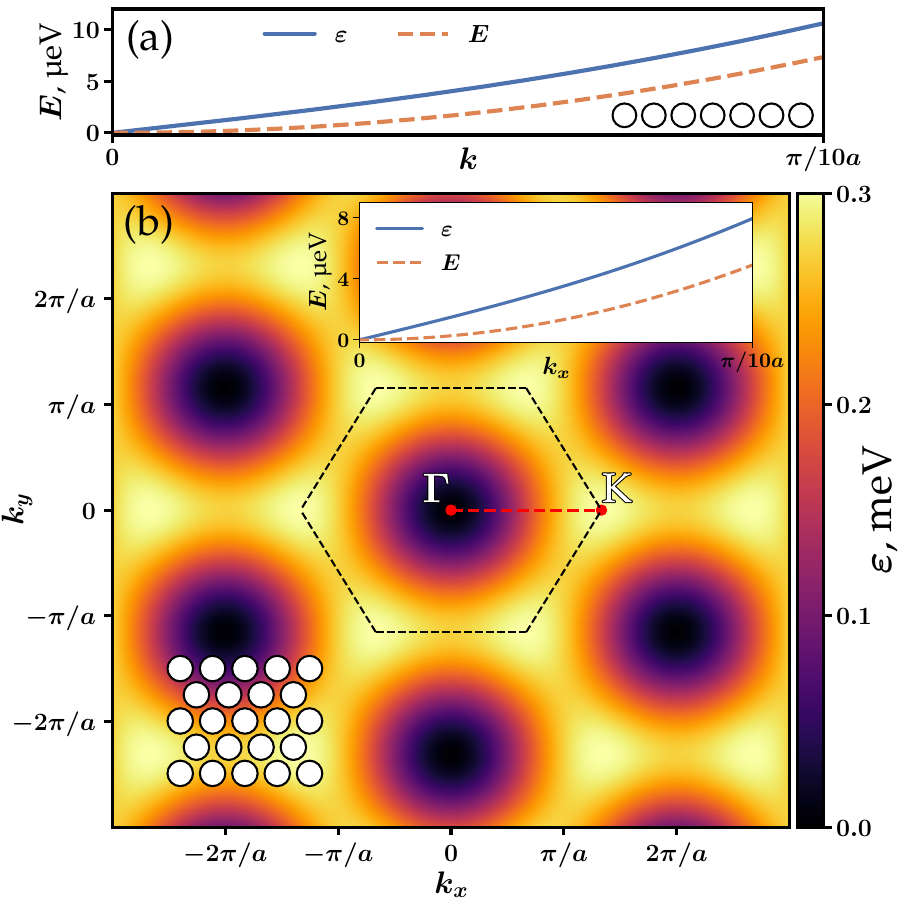}
    \caption{(a) The calculated single polariton dispersion (blue) and the characteristic dispersion of a Nambu-Goldstone mode (orange) of a bosonic condensate of exciton-polaritons in a linear chain of micropillars (shown schematically in the right bottom part) (b) The two-dimensional dispersion of Nambu-Goldstone modes in a triangular periodic array of exciton-polariton condensates schematically shown at the left bottom corner. The inset compares the dispersions of single polariton and Nambu-Goldstone modes plotted along the $\Gamma$-K trajectory in a Brillouin zone. In both panels, the chemical potential of the polariton condensate is taken to be $\mu=0.4$ \textmu eV.}
    \label{fig:E}
\end{figure}

In this work, we present a simple analytical model that may serve as an approach to such a microscopic theory. Within this model, we demonstrate that the KPZ scaling observed in the temporal and spatial decay of coherence in polariton condensates may stem directly from the stochastic dynamics of the Nambu-Goldstone modes. This conclusion is supported by numerical simulations that reproduce the experimentally observed KPZ scaling regimes and elucidate their dependence on system parameters such as the interaction strength and pump power.


\textit{The analytical model.}
It is well established in multiple experiments that arrays of coupled exciton-polariton condensates may spontaneously form extended coherent phase-locked modes~\cite{Lagoudakis_Couplings, Berloff_XY, Berloff_Ising, Lagoudakis_lattices, PRX_coupling, Topfer_lattices, cao2025polaritonxysimulatorsrevisited}. We shall consider one of such modes occupied by a Bose-Einstein condensate of exciton-polaritons in the mean-field approximation and describe it with a scalar space-dependent order parameter. In an infinite periodic lattice of condensates, the order parameter can be expanded on a plane wave basis as follows:
\begin{equation}
    \psi(\vb r, t) = \sqrt{n}\exp(i\vb k - i\omega(\vb k)t),
\end{equation}
where $n$ is the polariton density, $\vb k$ is the reciprocal lattice vector, $E(\vb k)=\hbar \omega(\vb k)$ is the mode energy.

In the vicinity of the bosonic condenation threshold, the integrated population of the excited states of the polariton condensate (Nambu-Goldstone modes) is comparable with the population of the condensate. In this regime, the fluctuating populations of the excited states which are characterized by random phases are expected to dominate the phase coherence dynamics in the system. We assume that the lowest energy $E_0=\hbar\omega_0$ polariton state characterized by the wavevector $\vb k_0$ is populated with the density $n_0$. The fluctuating phase of the polariton population comprising the ground state of the condensate and the Nambu-Goldstone modes can be obtained by all relevant wavefunctions with corresponding relative weights given by the Bogolyubov ansatz~\cite{Bogolyubov:1947zz}:
\begin{align} \label{eq:psi}
    \psi(\vb r, t)&=\exp(i\vb k_0\vb r-i\omega_0t)\bigg[\sqrt{n_0}\, + \nonumber \\  
    &+ \sum_{\vb k\neq 0}\sqrt{n(\vb k)}A_{\vb k}\exp(i\vb k\vb r - i \omega_B(\vb k)t)\, + \nonumber \\
    & + \sum_{\vb k\neq 0}\sqrt{n(\vb k)}B_{\vb k}\exp(-i\vb k\vb r + i \omega_B(\vb k)t)\bigg].
\end{align}
The Nambu-Goldstone modes are assumed to obey a conventional Bogolyubov dispersion relation:
\begin{equation}
    \varepsilon(\vb k) = \hbar\omega_B(\vb k) = \sqrt{E(\vb k)\qty(E(\vb k) + 2\mu)},
\end{equation}
where $\mu$ is the energy of polariton-polariton interactions playing role of the effective chemical potential.
Fig.~\ref{fig:E}(a) shows the dispersion curve characterizing Nambu-Goldstone modes in comparison with the bare polariton dispersion.
The relative amplitudes of Nambu-Goldstone modes $A_{\vb k}$ and $B_{\vb k}$ are governed by the Bose-Einstein distribution:
\begin{align}
    \abs{A_{\vb k}} &= \frac{1}{2}\qty(\frac{\varepsilon(\vb k)+\mu}{E(\vb k)}+1), \\
    \abs{B_{\vb k}} &= \frac{1}{2}\qty(\frac{\varepsilon(\vb k)+\mu}{E(\vb k)}-1),
\end{align}
We estimate the populations $n(k)$ of the excited states with use of the Bose–Einstein distribution:
\begin{equation}
    n_B(\vb k) = \qty(\exp(\frac{\varepsilon(\vb k)}{k_BT})-1)^{-1},
\end{equation}
where $T$ plays role of an effective temperature.


Now we are able to calculate the first-order correlation function of the system
\begin{equation}
    g^{(1)}(\Delta \vb r, \Delta t) = \frac{\expval{\psi^*(-\vb r,t_0)\psi(\vb r,t_0+\Delta t)}}{\sqrt{\expval{\psi(-\vb r,t_0)}^2}\sqrt{\expval{\psi(\vb r,t_0+\Delta t)}^2}},
\end{equation}
where $\Delta \mathbf r=2\mathbf r$, and $\langle ... \rangle$ denotes the ensemble averaging. Here the ensemble consists of all possible expansions of the form \eqref{eq:psi} that differ in the phases of the harmonics $A_{\vb k}$ and $B_{\vb k}$. Based on the ergodicity hypothesis, the ensemble averaging can be replaced by averaging over the initial time (time when our observation starts) $t_0$ for the fixed phases of $A_{\vb k}$ and $B_{\vb k}$:
\begin{align} \label{eq:g1}
    &g^{(1)}(\Delta \vb r, \Delta t)\propto \int \psi^*(-\vb r,t_0)\psi(\vb r,t_0+\Delta t) \dd t_0 \propto \nonumber\\
    &\ \propto n_0 + \sum_{\vb k\neq0}n(\vb k)\abs{A(\vb k)}^2 \exp{i\vb k\Delta \vb r-i\omega(\vb k)\Delta t}\, +\nonumber\\
    &\ +\sum_{\vb k\neq0}n(\vb k)\abs{B(\vb k)}^2 \exp{-i\vb k\Delta \vb r+i\omega(\vb k)\Delta t}
\end{align}

As Bose-Einstein condensation in one- and two-dimensional system is not possible for infinite systems~\cite{shlyapnikov2004lowdimensionaltrappedgases} we must take into account a finite linear size of the array of exciton-polariton condesnates $L$. The boundary conditions for finite-size polariton lattice leads to $k$-space discretization with step of $2\pi/L$. The summation in Eq.~\eqref{eq:g1} needs to be performed over the discrete set of polariton eigen-states in the first Brillouin zone.

In arrays of exciton-polariton condensates, the KPZ-regime manifests itself with the time- and space-depndencies of the phase $\theta$ described by the KPZ equation:
\begin{equation}
  \partial_t\theta(\vb r,t)=
      \nu\,\laplacian\theta
    +\frac{\lambda}{2}\bigl(\nabla\theta\bigr)^2
    +\xi(\vb r,t),
  \label{eq:KPZ}
\end{equation}
where $\nu$ and $\lambda$ are arbitrary coefficients and $\xi(\vb r, t)$ is the Gaussian white noise term. The corresponding KPZ-scaling can be revealed experimentally by measuring the first order coherence $g^{(1)}(\Delta \mathbf r, \Delta t)$~\cite{KPZ_1d}:
\begin{align}
    C(\Delta t)\equiv -2\log\abs{g^{(1)}(\Delta \vb r=0, \Delta t)}&\propto \Delta t^{2\beta}, \label{eq:Ct}\\
    C(\Delta\vb r)\equiv-2\log\abs{g^{(1)}(\Delta \vb r, \Delta t=0)}&\propto \abs{\Delta r}^{2\chi}, \label{eq:Cr}
\end{align}
where $\beta = 1/3$ and $\chi = 1/2$ corresponds to the one-dimensional case, and $\beta \approx 0.241$ and $\chi \approx 0.387$ corresponds to the two-dimensional case~\cite{KPZ}.


\textit{Numerical results.} 
To characterise the spatial and temporal phase fluctuations in the studied system we calculate the correlation function $g^{(1)}$ given by equation \eqref{eq:g1} for one- and two-dimensional lattices of exciton-polariton condensates.

\begin{figure}[h]
    \centering
    \includegraphics[width=0.99\linewidth]{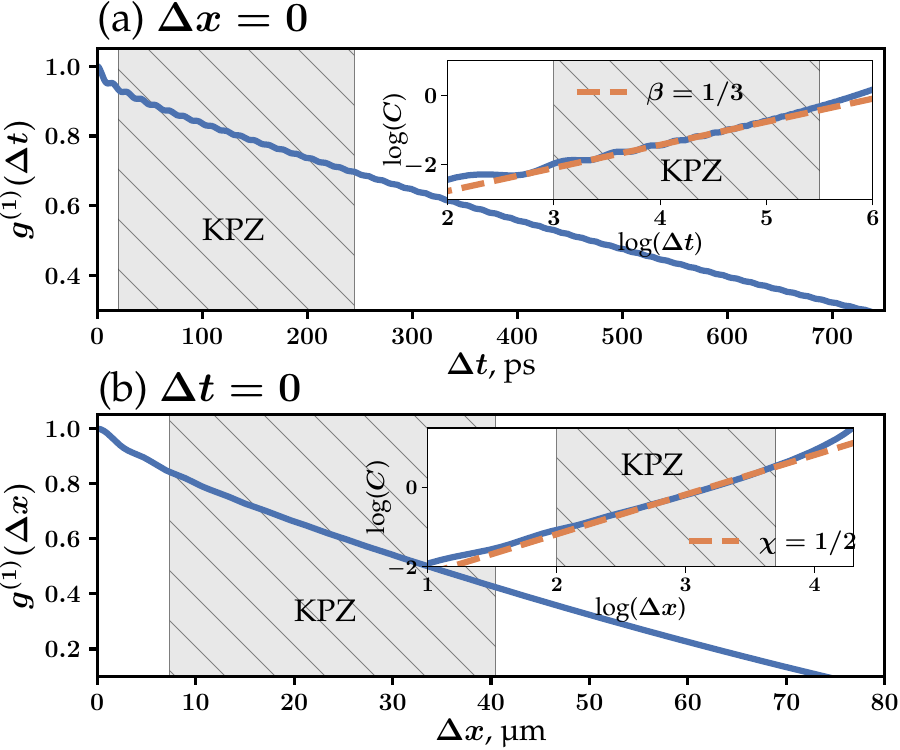}
    \caption{The time-dependent first-order correlation function $g^{(1)}(\Delta x=0, \Delta t)$ (a) and spatial correlation function $g^{(1)}(\Delta x, \Delta t=0)$ (b) calculated for a one-dimensional periodic array of exciton-polariton condensates. Insets are showing the same data in log-log scale fitted with KPZ dependencies \eqref{eq:Ct} and \eqref{eq:Cr}. KPZ regions are marked with hatched grey areas.}
    \label{fig:g1_1d}
\end{figure}
First, we consider a linear one-dimensional chain of polariton condensates. The energies $E(k)$ of phase-locked modes of such lattice are propotional to $\cos(ka)$, where $k$ is the reciprocal lattice vector and $a$ is the lattice constant~\cite{cao2025polaritonxysimulatorsrevisited}. The exact form of $E(k)$ depends on the lattice constant $a$ which defines coupling of neighbouring condensates. To set the macroscopically occupied extended condensate mode $E(k_0)=0$  we assume the following energy band structure for the periodic array of exciton-polariton condensates:
\begin{equation} \label{eq:E_1d}
    E(k) = \Delta E\cdot\qty(\frac{1 - \cos(k a)}2),
\end{equation}
where $\Delta E$ is the energy band width. The energy spectrum of Eq.~\eqref{eq:E_1d} for the polariton modes of a linear chain and the corresponding Bogolyubov dispersion are demonstrated in Fig.~\ref{fig:E}(a). The Bogolyubov spectrum features the linear dispersion of excitations in the longwave limit.

Fig.~\ref{fig:g1_1d} demonstrates temporal $g^{(1)}(\Delta x=0, \Delta t)$ (panel (a)) and spatial $g^{(1)}(\Delta x, \Delta t = 0)$ (panel (b)) correlation function calculated for one-dimensional chain. One can see that significant part of the dynamics correspond to the critical exponents $\beta=1/3$ and $\chi=1/2$ confirming the KPZ-scaling. 

\begin{figure}[]
    \centering
    \includegraphics[width=0.99\linewidth]{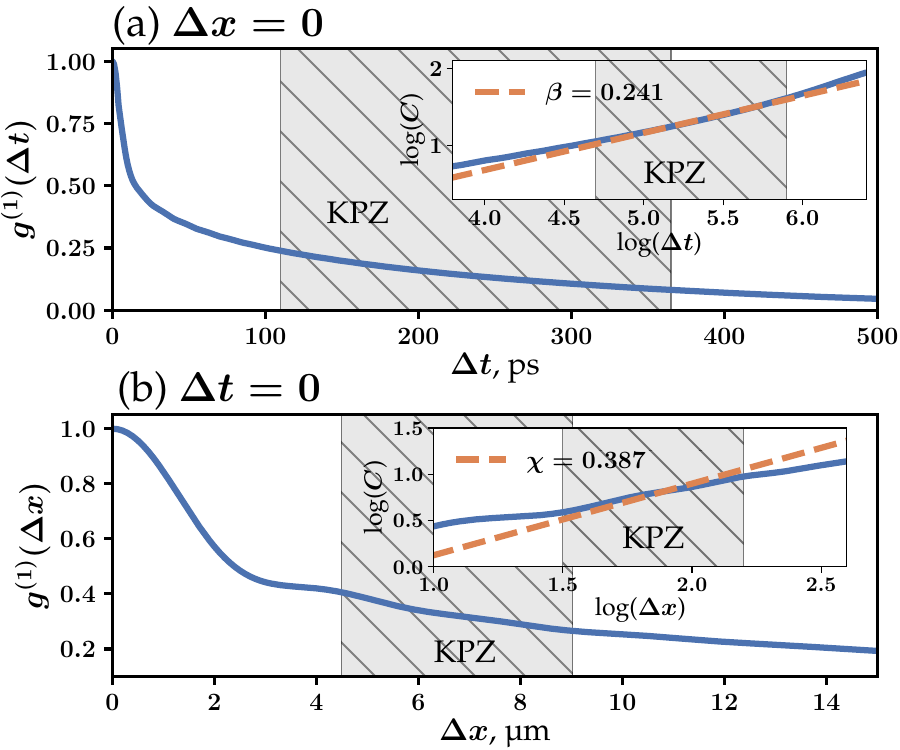}
    \caption{The time-dependent first-order correlation function $g^{(1)}(\Delta x=0, \Delta t)$ (a) and spatial correlation function $g^{(1)}(\Delta x, \Delta t=0)$ (b) calculated for a two-dimensional periodic array of exciton-polariton condensates. Insets show the same data in log-log scale fitted with analytic KPZ dependencies \eqref{eq:Ct} and \eqref{eq:Cr}. KPZ-regions are selected as hatched grey areas.}
    \label{fig:g1_2d}
\end{figure}

To reveal KPZ-scaling in two dimensions, we consider a triangular lattice of exciton-polariton condensates as Fig.~\ref{fig:E}(b) illustrates. Such triangular lattice was recently used to experementally observe the KPZ scaling by Widmann et. al.~\cite{KPZ_Hofling}. In this case, the dispersion of phase-locked lattice modes is governed by the geometric structure factor ${\gamma(\vb k)=\cos(k_xa)+2\cos(k_xa/2)\cos(\sqrt{3}k_xa/2)}$~\cite{cao2025polaritonxysimulatorsrevisited}. In the same way as in the one-dimensional case, we assume the conventional 2D energy band dispersion:
\begin{align}
    E(\vb k) &= \frac{2\Delta E}{9} \cdot \bigg[3-\cos(k_xa)\,+ \nonumber\\ &+2\cos(\frac{k_xa}2)\cos(\frac{\sqrt{3}k_xa}2)\bigg].
\end{align}

 Fig.~\ref{fig:g1_2d} shows the calculated correlation function $g^{(1)}$  in the considered 2D lattice of polariton condensates. As well as in the one-dimensional lattice, here we find regions with temporal ($\beta\approx0.241$) and spatial ($\chi\approx0.387$) critical exponents characteristic of the KPZ-scaling regime. The typical scales of temporal and special fluctuation dynamics are given by the correlation time and correlation length of the system, in a remarkable similarity with the experimental results ~\cite{KPZ_Hofling}. 

 The oscillations of the correlation function and the limited size of KPZ regions are caused by the discretization of Brillouin zone due to the finite size of the system assumed in the model. These oscillations are not resolved in the experiment. Furthermore, the deviation from KPZ-scaling for smaller values of $\Delta t$ and $\Delta \vb r$ are due to the saturation of $g^{(1)}$: at the short timescale and the distances smaller than the lattice constant phase fluctuations caused by the excited states of the extended condensate mode are insignificant.
 Clearly, the simple analytical model considered here may not yield an exact agreement with the experimental data. However, it captures the main trends and privides an important information on the microscopic mechanism that is behind KPZ-scaling in arrays of polariton condensates.

\begin{figure}[h]
    \centering
    \includegraphics[width=0.99\linewidth]{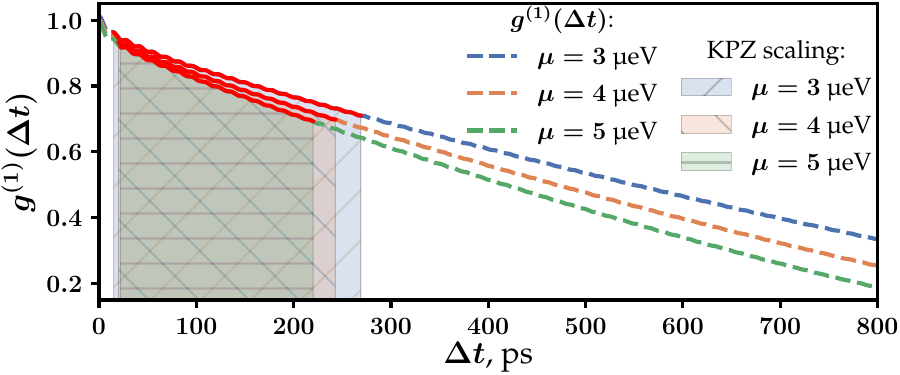}
    \caption{The time-dependent first-order correlation function $g^{(1)}(\Delta x=0, \Delta t)$ of a one-dimensional chain of exciton-polariton condensates calculated for different interaction energies $\mu$: 3 \textmu eV (blue), 4 \textmu eV (orange) and 5 \textmu eV (green). KPZ-regions are marked with red lines and hatched areas of the corresponding color.}
    \label{fig:g1_mu}
\end{figure}

In the experiments, KPZ-scaling in polariton condensates has been detected at the pump intensities in the close vicinity of the bosonic condensation threshold. This is fully consistent with our analysis. As the pump power increases, the condensate density rises, and, consequently, the phase dynamics becomes more and more dependent on the contribution of the condensate itself, while the role of its excited states becomes less important. In addition, the effective chemical potential increases $\mu\propto N_\text{pol}$ which affects the slope of the linear part of the dispersion of Nambu-Goldstone modes. Consequently, the overall population of these modes decreases which also leads to the reduction of their impact on the fluctuation dynamics. Fig.~\ref{fig:g1_mu} presents the temporal correlation functions $g^{(1)}(\Delta x=0, \Delta t)$ calculated for a one-dimensional condensate for different values of $\mu$.

These calculations show that as the pump power increases, the time interval over which KPZ scaling prevails decreases. As a result, the robust KPZ-scaling can be observed only at the relatively low pump power intensity where the integrated population of Nambu-Goldstone modes is comparable with the condensate population.

The parameters used in numerical simulations: ${\Delta E=0.3}$~meV, ${a=2}$~\textmu m, ${T=300}$~K, ${L=200}$~\textmu m, ${\mu=4}$~\textmu eV (for Fig.~\ref{fig:E}-\ref{fig:g1_2d}).


\textit{Conclusion.} Within the analytical toy-model, we have been able to conclude on a likely microscopi mechanism for the experimentally observed KPZ-scaling in one-dimensional and two-dimensional arrays of exciton-polariton condensates. We suggest that the reason for such scaling is strongly linked to the spontaneous symmetry breaking in a driven-dissipative polaritonic system resulting in the fluctuating populations of the Nambu-Goldstone modes. The pump power dependence of the fluctuation dynamics in arrays of polariton condensates indicated that the KPZ-scaling dominates in the low-power regime where the integrated population of the Nambu-Goldstone modes is comparable with the population of the condensate. At the higher pump power, the pupulation of the condensate increases, while the mean population of Nambu-Goldstone modes decreases. This is why the KPZ-dynamics is replaced by an equilibrium Berezinskii-Kosterlitz-Thouless dynamics~\cite{KPZ_Hofling}.  

Our work not only offers an explanation of the possible origin of KPZ-universality in polaritonic systems but also opens a pathway for controlling the scaling behavior in any system containing bosonic condensates by manipulating their Nambu-Goldstone modes. This offers a tool which may prove precious for application in the development of quantum light sources with tailored photon correlation properties.

\bibliography{Reference}

\end{document}